\documentclass[11pt,twoside]{article}
\usepackage{asp2014}

\aspSuppressVolSlug
\resetcounters

\begin{document}

\title{Taplint, the TAP Service Validator}

\author{Mark~Taylor}
\affil{H.~H.~Wills Physics Laboratory, Tyndall Avenue,
       University of Bristol, UK;
       \email{m.b.taylor@bristol.ac.uk}}

\paperauthor{Mark~Taylor}{m.b.taylor@bristol.ac.uk}{0000-0002-4209-1479}{University of Bristol}{School of Physics}{Bristol}{Bristol}{BS8 1TL}{U.K.}



  
\begin{abstract}
TAP, the Table Access Protocol, is a widely used Virtual Observatory
specification
allowing client software to interact with remote database services in
a standardised way.
This paper presents {\tt taplint}, a tool for assessing the compliance
of deployed TAP services with the the dozen or so formal
specifications that form the TAP protocol stack.
We provide an overview of its capabilities and operation,
and the context within which it is used to
improve robustness of data services.
\end{abstract}

\section{Introduction}

TAP, the Table Access Protocol \citep{2019ivoa.spec.0927D}
is a Virtual Observatory protocol suite
allowing client software to interact with remote database services
in a standardised way, including acquiring rich metadata and submitting
simple or complex queries in an SQL-like language.
It is a VO success story,
underlying much current access to astronomy data archives,
for instance providing the primary access to ESA's Gaia Archive
\citep{2016A&A...595A...2G}.
The protocol stack involved however is quite complex,
involving a dozen or so separate IVOA specifications,
and implementors face many opportunities to make mistakes.

The International Virtual Observatory Alliance (IVOA)
is the standards body within which TAP has been developed.
The IVOA {\em Document Standards} specification \citep{2017ivoa.spec.0517G},
which defines how protocol standards are developed and adopted,
requires that a validator tool, as well as interoperable implementations,
must be available before a standard can be endorsed.
This requirement is in place partly as a check of the written specification,
since implementing a validator is a good way to pick up ambiguities and
inconsistencies in the text,
and partly as an aid to both implementors and users of the protocols.
Running validation tests on a service, ideally as part of the
development and deployment process, can identify errors and issues
that could otherwise cause trouble for service users.

Within this context, the {\tt taplint} tool has been developed as
a validator for deployed TAP services.
Since TAP is closely related to a number of other IVOA and some non-IVOA
specifications, {\tt taplint} serves as a validator for a number of
standards beyond the TAP document itself.

\section{Usage and Availability}

{\tt taplint} is provided as part of STILTS \citep{2006ASPC..351..666T}, 
which is a command-line package for manipulating catalogues and other tables,
written in Java.
To run it, the only requirement is a Java Runtime Environment
(Java 8 or later), and the {\tt stilts.jar} file
that can be downloaded from the package
web page\footnote{\url{http://www.starlink.ac.uk/stilts/stilts.jar}}.
The validator can then be run by supplying on the command line the base URL
of the TAP service to test:
\begin{verbatim}
   java -jar stilts.jar taplint tapurl=http://example.com/tap
\end{verbatim}
Additional command-line parameters may optionally be supplied
to control which testing stages are run, 
the form of the output, and some details of service interaction.

When run, the tool writes a series of reports to standard output.
By default the output is line-oriented, {\tt grep}-friendly,
and intended for human consumption (though JSON output is also
available).
Each report line is of the form:
\begin{verbatim}
   T-SSS-MMMMxN aaaaa...
\end{verbatim}
where the parts have the following meanings:
\begin{description}
\item[{\tt T}:]
      report type,
      one of {\tt I}(nfo), {\tt W}(arning), {\tt E}(rror), {\tt F}(ailure)
         and {\tt S}(ummary).
      Info gives information about the test being performed,
      Warning reports behaviour that is questionable or
      not recommended by standards, and
      Error reports behaviour contrary to standards
\item[{\tt SSS}:]
      3-character stage identifier,
      indicating which part of the test sequence is in progress
\item[{\tt MMMM}:]
      4-character unique code identifying the test being run; 
      there are currently 200$+$ of these
\item[{\tt xN}:]
      repeat count for multiple occurrences of reports with the same
      {\tt SSS-MMMM}
\item[{\tt aaaa...}:]
      informative message
\end{description}
An example report would be (presented here on two lines for readability):
\begin{verbatim}
   E-OBS-CUTP-7 Wrong Utype in ObsCore column access_format:
                Access.format != obscore:Access.Format
\end{verbatim}

To limit the quantity of output
no more than a fixed maximum, by default 10, reports with the same
{\tt SSS-MMMM} will be written.
This prevents the output file becoming unwieldy when
the same or a similar problem is encountered in multiple tests.

Considerable efforts are made to explain the nature of the issues
reported in the message text.
However, failures can be subtle, and moreover the validator,
like other software, may contain bugs, so users are encouraged
to contact the author with queries about {\tt taplint}
output or behaviour.

A typical run might take three minutes and generate 200 lines of output,
but these values can vary widely depending on the service.
Full details of the tool's operation are given in the STILTS
documentation\footnote{\url{http://www.starlink.ac.uk/stilts/sun256/taplint.html}}.

\section{Standards Covered}

The TAP standard itself defines how clients can interrogate metadata,
pose queries, and retrieve results from remote database services.
However it does this with reference to many other IVOA (and some non-IVOA)
standards concerning serialization, service interaction,
metadata encoding, capability declaration and more;
there are also further standards defining domain-specific data models
that sit on top of TAP.  {\tt taplint} understands most of these,
and runs specific tests that a target service is behaving as prescribed.

Behaviour defined by the following IVOA specifications
is tested to a greater or lesser extent:
TAP, VOTable, UWS, VODataService, ADQL, VOResource, VOSI, TAPRegExt,
DALI, ObsCore, ObsLocTAP, EPN-TAP, UCD, VOUnits, SoftID and SSO.
These documents are available from the IVOA {\em Documents and Standards\/}
web page\footnote{\url{https://www.ivoa.net/documents/}},
and are the result of much collaborative work over many years
between IVOA members.

\section{Tests Performed}

{\tt taplint} runs a battery of test queries against the TAP service,
posing as a client and checking that the response in each case
is as mandated by the relevant specifications.
It tries to test as many aspects of the required behaviour as it can,
especially in the areas of metadata provision,
output formats, job submission and compliance with data models.

The tests are grouped in a sequence of {\em stages\/};
by default all those applicable to the target service are performed,
but the list of stages can be restricted if only certain aspects
are of interest for a given run.
Tests include validation of output documents against relevant XML schemas,
careful checking of VOTable output against all the requirements of the
VOTable specification, submitting ADQL queries in various modes,
checking that table metadata is standards-compliant and consistent
between various ways of obtaining it,
testing asynchronous job submission behaviour, 
and testing that declared metadata is in line with
data models applying to served data.

{\tt taplint} tries its best to test that all the requirements
and recommendations of the standards it knows about
are implemented as specified in those standards.
It should be noted however that no validator can supply all possible
inputs to a complex service like TAP,
so just because {\tt taplint} does not identify errors in a service
does not guarantee that none are present.

\section{Discussion}

{\tt taplint} provides a way to identify flaws in running TAP services
during development and operations,
and can thus contribute to the efforts of service providers in
improving the experience for their users,
since broken or partially working services can be a major
cause of user frustration.
It has been used at many data centers for this purpose,
including by ASDC, CADC, CDS, CfA, ESA, ESO, GAVO, IPAC, MAST and NASA,
as well as providing the TAP-specific component of the bulk validation
services run by ESA, PADC and NASA to assess overall VO operational
status \citep[see {e.g.}][]{2020ASPC..522..339S}.
Data providers developing or operating a TAP service are encouraged
to consider running {\tt taplint} to check its compliance with the standards.

The VO standards landscape is in constant flux,
and {\tt taplint} requires frequent updates
in response to new and updated specifications,
changing or contested interpretations of existing documents
and user feedback.
Recent improvements (STILTS v3.4-2)
include a new UCD and Unit validation stage,
validation of the data and metadata
specified by the new EPN-TAP and ObsLocTAP data models,
better stage selection control,
checking service VO component identification,
improved reporting of VOTable issues, and
validation of content declared by xtype.

Writing validation tools like {\tt taplint} is
(like much of the work in the VO) time-consuming and not very glamorous,
but they can provide an important contribution to the robustness
of the service ecosystem, as well as feeding back to improve the
standards definition process.

\acknowledgements This work has been supported by STFC and GAVO. Development of the software has benefitted from much input from its users and collaborative discussions with VO experts.

\bibliography{X7-008}


\end{document}